# Un marco conceptual para la generación de requerimientos de software de calidad

Mauro José Pacchiotti[1] , Mariel Ale[1] y Luciana Ballejos[1]

[1] Centro de I+D de Ingeniería en Sistemas de Información
UTN-FRSF
{mpacchiotti, lballejos, male}@frsf.utn.edu.ar

**Resumen.** Los requerimientos expresados en lenguaje natural son un artefacto indispensable en el proceso de desarrollo de software, ya que pueden ser comprendidos por todas las partes interesadas. Sin embargo, su ambigüedad representa un desafío persistente. Para abordar esta problemática, organizaciones como IEEE e INCOSE publican guías para escribir requerimientos con reglas que ayudan en la tarea. Por otro lado, las metodologías ágiles tienen patrones y estructuras para escribir con lenguaje natural las necesidades de los involucrados, tratando de restringir el lenguaje para evitar la ambigüedad. Sin embargo, la diferencia de conocimiento entre los involucrados sobre los requerimientos y la forma correcta de expresarlos, dificulta aún más la tarea de especificación. En los últimos años, los grandes modelos lingüísticos (LLM, por sus siglas en inglés) han emergido para mejorar las tareas de procesamiento de lenguaje natural, estos son arquitecturas basadas en modelos profundos que emulan mecanismos de atención similares a los humanos. Este trabajo tiene como objetivo probar el poder demostrado por los LLM en este dominio. El propósito es utilizar estos modelos para mejorar la calidad de los requerimientos de software expresados en lenguaje natural, ayudando a los analistas en las tareas de especificación de requerimientos. Se detallan el marco propuesto, la arquitectura, los componentes clave y las interacciones entre ellos. Además, se desarrolla una prueba conceptual de la propuesta para evaluar su utilidad. Por último, se discuten el potencial y las limitaciones del marco de trabajo, así como las direcciones futuras para continuar su validación y refinamiento.

**Palabras clave:** Ingeniería de Requerimientos, Calidad de Requerimientos, Grandes Modelos Lingüísticos.

## 1 Introducción

El Proceso de Ingeniería de Requerimientos (IR) es esencial en el desarrollo de software. El objetivo de este proceso es garantizar que se incorporan todas las necesidades, limitaciones y exigencias de calidad del cliente, y que el software a producir las satisface. Los requerimientos generados en esta fase proceden de varios tipos de fuentes: distintos documentos del dominio, stakeholders y sistemas heredados, si existen (Wiegers et al., 2013). Dos de estas fuentes, los documentos y los stakeholders, tienen como característica principal el uso del lenguaje natural. Por un lado, los documentos están escritos en lenguaje natural y por otro, el lenguaje natural es el medio de comunicación con las partes interesadas y la herramienta fundamental para lograr la validación de los requerimientos.

El lenguaje natural tiene características no deseadas, tales como la ambigüedad, que dificultan la redacción del documento resultante del proceso de IR, llamado Especificación de Requerimientos de Software (ERS). Esta especificación contiene todas las necesidades, restricciones y demandas de calidad obtenidas de las partes interesadas y extraídas de los documentos y sistemas heredados. Un uso incorrecto del lenguaje natural en el proceso y en la creación este documento, puede dar lugar a malentendidos u omisiones que dificulten el desarrollo del software en etapas posteriores.

INCOSE publica y mantiene una guía para escribir requerimientos (GtWR Homepage, 2025) que recomienda un conjunto de características y reglas para ayudar a los analistas en el proceso de especificación de requerimientos. En esta guía, se enumera un conjunto de características no deseadas, y la ambigüedad es la más común de ellas. La ambigüedad existe con frecuencia en las expresiones del lenguaje natural en varios sentidos, y es necesario eliminarla de las expresiones de requerimientos, porque éstas necesitan tener un significado único.

En los últimos tiempos, con el crecimiento de la inteligencia artificial, se han propuesto nuevos modelos profundos que han demostrado un gran rendimiento en el procesamiento del lenguaje natural. Estos modelos, llamados LLM, se basan en un modelo base propuesto en el trabajo *Attention is all that you need* (Vaswani et al., 2017), llamado Transformer. A partir del Transformer han surgido varios modelos con variantes de su arquitectura, algunos de los cuales han demostrado mejoras significativas en tareas de lenguaje natural, tales como la generación de resúmenes (Liu et al., 2024), traducciones (Volk et al., 2024) y respuestas a preguntas (Robinson et al., 2022), entre otras.

Este artículo pretende incorporar el uso de LLM para ayudar al analista en tareas de especificación de requerimientos, basándose en las grandes cualidades de estos modelos para mejorar el procesamiento de lenguaje natural. Se propone un marco de trabajo para interactuar con los modelos en varias tareas dentro de un proceso que pretende mejorar la calidad de los requerimientos de software especificados.

El resto de este trabajo se organiza como sigue: la Sección II presenta los fundamentos, la Sección III describe el marco propuesto, la Sección IV desarrolla una prueba conceptual de la propuesta y evalúa los resultados, y la Sección V discute los resultados, expone las conclusiones y propone los trabajos futuros.

## 2 Marco Teórico

### 2.1 Proceso de IR

La IR es una rama de la Ingeniería del Software (IS) que se ocupa de la gestión de los servicios y restricciones de los sistemas de software (Nuseibeh et al., 2000). El proceso de IR puede dividirse en cuatro actividades principales: estudio de factibilidad, obtención y análisis, especificación y validación de requerimientos, según Sommerville (Sommerville, 2016). Estas actividades no se ordenan secuencialmente, sino que pueden realizarse de forma iterativa a lo largo del proceso y a medida que cambian los requerimientos o se descubren nuevos. El marco propuesto en este trabajo pretende asistir a los analistas principalmente en las actividades de obtención, análisis y especificación de requerimientos.

### 2.2 Grandes Modelos Lingüísticos

A partir del trabajo *Attention is all that you need* (Vaswani et al., 2017), se propuso el Transformer, un modelo de aprendizaje profundo con una arquitectura dividida en dos estructuras, el codificador y el decodificador, basado en el mecanismo de atención multicabeza. Mientras que los modelos recurrentes procesan el texto secuencialmente, el Transformer utiliza un mecanismo de atención que le permite considerar todas las palabras de una frase simultáneamente. Este mecanismo permite a los modelos lingüísticos destacar las partes más relevantes de un texto de entrada para comprender mejor su significado y contexto. Así, el modelo puede captar las relaciones de largo alcance entre las palabras, lo que se traduce en un mejor rendimiento en diversas tareas de lenguaje natural.

Desde el Transformer, se siguieron desarrollando modelos basados en su estructura, tales como BERT (Bidirectional Encoder Representations from Transformers) (Devlin et al., 2018), GPT (Generative Pretraining Transformer) (Brown et al., 2020) y T5 (Text-to-Text Transfer Transformer) (Raffel et al., 2019), que se entrenaron con conjuntos de datos aún mayores, compuestos por texto y código. Estos modelos destacan por su capacidad para generar texto de alta calidad y realizar diversas tareas de PLN. Posteriormente, el uso de LLM se extiende aún más con la aparición de ChatGPT (ChatGPT Homepage, 2025) en 2022, una aplicación de la empresa OpenAI que permite utilizar IA generativa a través de una interfaz de chat basada en el modelo GPT para generar respuestas.

### 2.3 Prompt Engineering

La interacción con LLM generativos propone un nuevo paradigma en el que se puede interactuar con un modelo con expresiones en lenguaje natural. En estos modelos de lenguaje, la entrada del usuario se denomina *prompt*, que es una instrucción o conjunto de palabras diseñadas para iniciar y guiar la respuesta del modelo. Es importante la forma en que se genera esta expresión, ya que el modelo generará la salida a partir de la misma. Las buenas prácticas recomiendan elementos que un prompt puede contener en busca de lograr la respuesta esperada (Tabla 1).

**Tabla 1.** Elementos recomendados en un prompt.

| Elemento | Descripción |
| --- | --- |
| Instrucción | Tarea específica que se desea |
| Contexto | Información adicional que puede orientar al modelo y completar la respuesta. |
| Entrada | La entrada sobre la que se desea la acción. |
| Salida | Formato que se desea para la respuesta del modelo. |

Los cuatro elementos en un prompt no son obligatorios. Incluirlos o no, depende de la necesidad expresada. Además, en la entrada que se le proporciona al modelo se pueden incluir ejemplos de la tarea que esperamos del mismo. Dependiendo de la cantidad de ejemplos que se le proporcionen al modelo, la ingeniería de prompts se puede categorizar en tres enfoques principales:

• Zero-shot: La entrada no incluye ejemplos de la tarea a realizar.
• One-shot: La entrada incluye un solo ejemplo de la tarea que se desea realizar.
• Few-shots: La entrada incluye varios ejemplos de la tarea. El few-shots es el enfoque más poderoso, ya que el modelo puede aprender patrones y generalizaciones a partir de múltiples ejemplos.

### 2.4 Generación Aumentada por Recuperación

A pesar del alto rendimiento de los LLM, su conocimiento se limita a los datos con los que fueron entrenados, que pueden no cubrir información contextual y estar desactualizados. Esta limitación reduce su capacidad para proporcionar información precisa y específica del dominio, independientemente de otros problemas, especialmente en campos especializados y dinámicos. Los sistemas de generación aumentada por recuperación (RAG, por su sigla en inglés Retrieval-augmented Generation) (Figura 1) se utilizan para superar esta limitación, mejorando las capacidades de los LLM mediante la recuperación de información relevante de fuentes externas durante el proceso de generación.

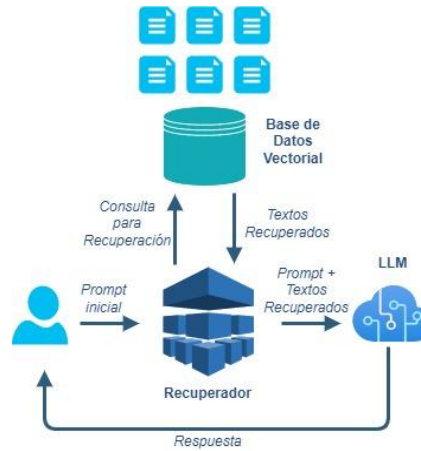

**Fig. 1.** Proceso de RAG.

## 3 Marco de trabajo propuesto

Esta propuesta se basa en interacciones entre requerimientos, fuentes de información y modelos con el objetivo de mejorar la calidad de los requerimientos. A continuación, se describe la arquitectura (Figura 2) y los distintos componentes que pueden implementarse dependiendo del dominio, alcance y tipo de problema a resolver

### 3.1 Arquitectura

La arquitectura propuesta da soporte al proceso que tiene como objetivo mejorar la calidad de los requerimientos. Este proceso comienza con un primer modelo que evalúa la expresión de requerimiento. Esta evaluación puede implementarse con distintos tipos de LLM o ensambles, dependiendo de qué tan amplias se pretendan la evaluación y la salida. Puede utilizarse desde un simple codificador como parte de un clasificador binario, hasta un modelo generativo que devuelva una salida detallada describiendo la calidad del requerimiento. Es importante definir y documentar los criterios y modelo o modelos que se utilizarán en esta primera etapa, a fin de sustentar la decisión posterior de mejorar o no la expresión.

El segundo modelo se ejecuta si, a partir de la salida del primero, se concluye que la calidad de la expresión del requerimiento no es suficiente. Para este caso, el segundo modelo analiza la expresión del requerimiento y la salida obtenida en la primera tarea, con el fin de obtener preguntas que mejoren su calidad.

Las preguntas generadas en la segunda tarea tienen como objetivo obtener respuestas que con su contenido aporten a mejorar la expresión del requerimiento, ya sea completando información o despejando dudas que disminuyan la ambigüedad de la expresión. Dependiendo del dominio, estas preguntas pueden utilizarse con dos tipos de fuentes. Por un lado, puede consultarse directamente a stakeholders y, por otro, pueden utilizarse con el tercer modelo propuesto y una fuente documental, si se dispone de ésta.

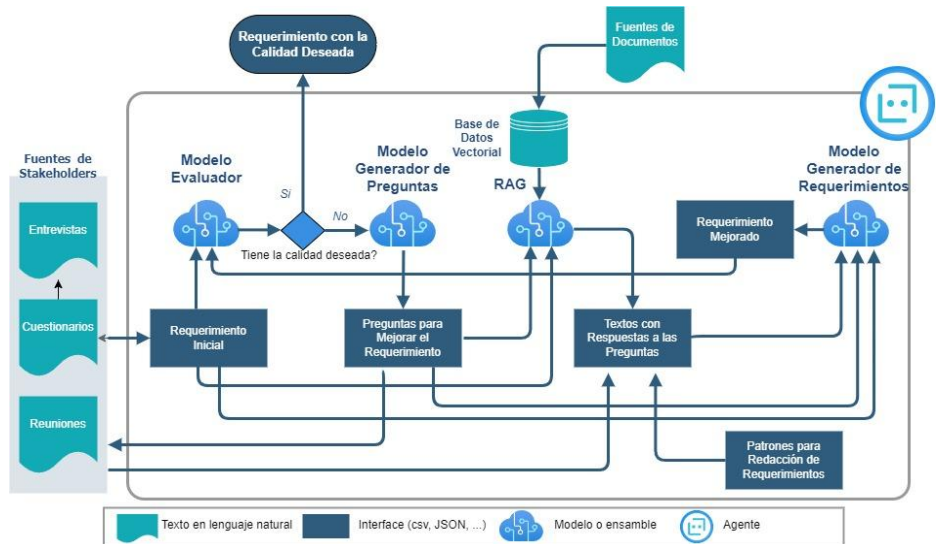

**Fig. 2.** Arquitectura de la propuesta.

El tercer modelo es una arquitectura RAG que toma como entradas un conjunto de preguntas sobre una expresión de requerimiento y busca en una base de datos vectorizada, con información de contexto, las posibles respuestas a estas preguntas. La utilización de este modelo depende de la existencia de documentación.

Un cuarto modelo recibe como entradas: el requerimiento original, un conjunto de preguntas y respuestas, fragmentos de contexto y patrones para redacción de requerimientos -si se dispone de éstos- para generar un nuevo requerimiento mejorado.
Finalmente, esta nueva expresión generada vuelve a ser evaluada por el primer modelo, a fin de verificar que cumple con la calidad esperada. De no ser así puede comenzarse una nueva iteración del proceso descripto.

### 3.2    Modelos

Los modelos implementados en el marco pueden variar dependiendo de las características del dominio y la tarea a desarrollar. Estos pueden ser modelos generativos preentrenados, ajustados con datos específicos, modelos clasificadores y hasta ensambles de varios modelos orquestados por un agente. Además, con el avance de estas tecnologías, los modelos pueden reemplazarse por otros con mejor performance.

### 3.3    Interfaces

Si la implementación del marco pretende automatizar las interacciones entre modelos, es necesario definir interfaces que estos puedan acceder para llevar adelante el proceso. Estas interfaces pueden ser desde simples documentos en texto plano, hasta bases de datos documentales.

### 3.4    Agente

La propuesta no exige el uso de un agente orquestador. Éste es necesario si se pretende implementar de forma automática parte o todo el proceso. El agente es el que orquesta la ejecución del proceso, interpretando interfaces, tomando decisiones y llamando a modelos dependiendo de las salidas obtenidas y los objetivos planteados en la implementación. Esto deja claro que, en caso de una implementación totalmente automática, el diseño del agente es un punto clave a resolver.

### 3.5    Ejemplos de Implementación

A continuación, se muestran dos posibles ejemplos de implementación del marco con distintas adaptaciones de la arquitectura propuesta.
**Ejemplo A (Figura 3).** En esta implementación se utilizan tres modelos generativos. (a) es modelo preentrenado, ajustado, que evalúa la expresión, (b) es un modelo preentrenado, ajustado, que genera preguntas para mejorar la expresión y (c) es un modelo preentrenado y utilizado con una técnica de few-shot prompting que genera los requerimientos de acuerdo con una plantilla. En esta implementación no se utiliza un agente y es el usuario el que cumple este rol, primero decide si el requerimiento tiene la calidad deseada de acuerdo con la salida del modelo (a), y luego necesita recolectar de los stakeholders las respuestas a las preguntas generadas por el modelo (b), para finalmente darles el formato adecuado para que sean utilizadas en la última etapa.

**Ejemplo B (Figura 4).** Esta propuesta de implementación utiliza primero un clasificador (a) que evalúa desde distintas perspectivas la expresión para evaluar si ésta tiene la calidad deseada. Si no es así, un modelo generativo preentrenado y ajustado (b) genera las preguntas para mejorar la expresión, luego una arquitectura RAG provee las respuestas a las preguntas utilizando documentos de estándares y transcripciones de entrevistas y reuniones. Finalmente, un modelo generativo preentrenado utilizado con una técnica de few-shot prompting genera los requerimientos de acuerdo con una plantilla. En este caso, el proceso es totalmente automático por lo que se necesita un agente que orqueste el funcionamiento de la implementación.

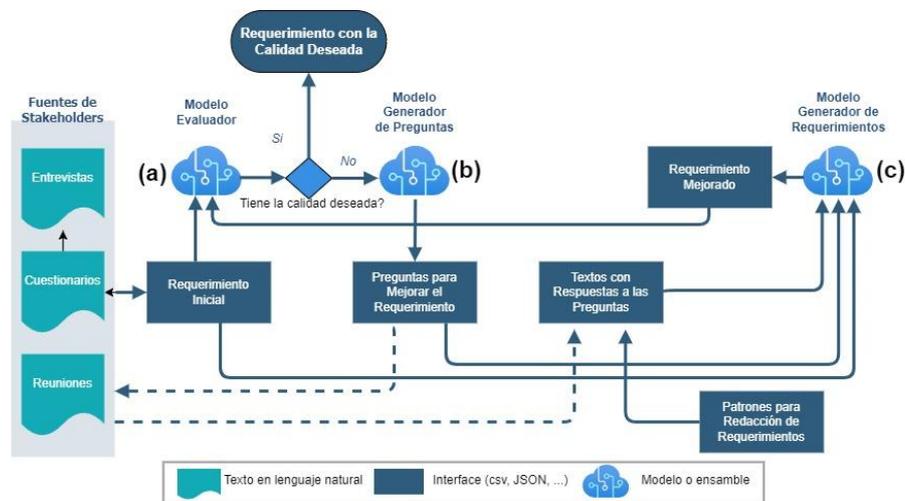

**Fig. 3.** Implementación del ejemplo A.

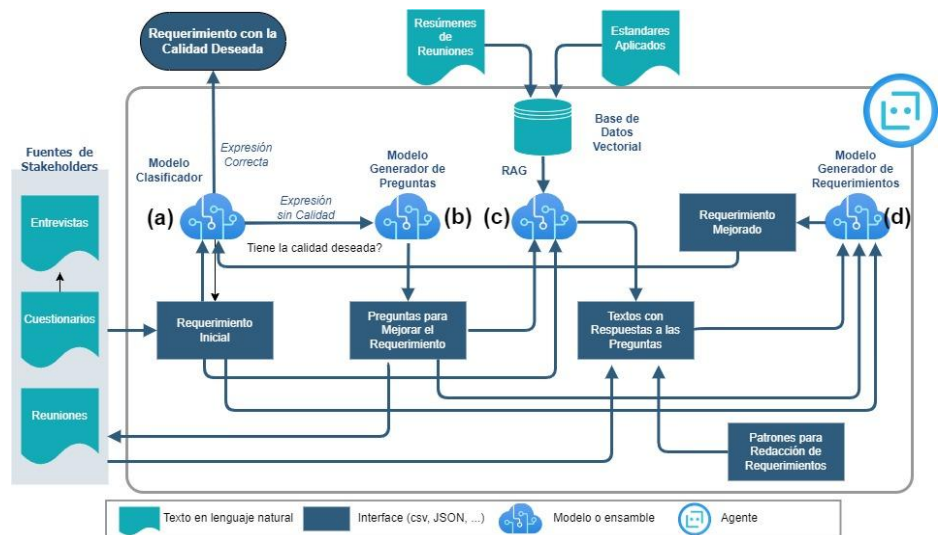

**Fig. 4.** Implementación del ejemplo B.

## 4   Prueba conceptual

### 4.1   Modelo utilizado

Es conocida la utilización de distintos asistentes, tales como Gemini (Gemini Homepage, 2025), Copilot (Copilot Homepage, 2025) o ChatGPT (ChatGPT Homepage, 2025), basados en LLM para dar soporte en infinidad de tareas, y no escapan a esta tendencia las actividades desarrolladas en el proceso de IR. Es por este motivo que las pruebas iniciales de esta propuesta se desarrollan utilizando un asistente bien conocido, ChatGPT versión 4o mini, aplicando técnicas de Prompt Engineering. Cabe aclarar que los resultados de las pruebas no son reproducibles con exactitud, debido a la naturaleza no determinística de las salidas que ofrecen estos modelos, aunque se repitieron algunas de las mismas obteniéndose resultados muy similares. Por último, el idioma elegido para el desarrollo de las pruebas es el inglés.

### 4.2   Metodología

Para el desarrollo de las pruebas se utilizan tres requerimientos (Tabla 2) que fueron escritos para estas pruebas, buscando que fueran expresiones sencillas de comprender y que representen características o requerimientos funcionales y no funcionales. Estas pruebas consisten en, inicialmente, utilizar el asistente propuesto, ChatGPT, mediante una consulta genérica para que sugiera mejoras a cada requerimiento por separado. Luego, para cada requerimiento, se utiliza una implementación simplificada del marco propuesto con el mismo objetivo. Al finalizar las pruebas, se pretende comparar las diferencias en los resultados obtenidos, con el fin de poder tener una primera evaluación de la propuesta que permita mejorarla y avanzar con pruebas a mayor escala.

**Tabla 2.** Expresiones de Requerimientos utilizadas en las pruebas iniciales.

| #  | Expresión |
|----|-----------|
| R1 | The system must allow the inventory manager to generate a list of missing products. |
| R2 | The system will have a user-friendly interface and support all common browsers. |
| R3 | A customer can cancel an order if he has not yet received it. |

### 4.3   Evaluación genérica de las expresiones

Las tres expresiones de requerimientos fueron utilizadas para realizar una consulta genérica a un modelo que tiene como objetivo mejorar la calidad del requerimiento. Además, el prompt utilizado pretende obtener en la respuesta del modelo recomendaciones para mejorar la expresión y una nueva expresión mejorada.
Para la redacción de la consulta se siguieron las técnicas recomendadas de prompt engineering en busca de una mejor respuesta. Como se puede observar en la Figura 5, inicialmente la expresión llama a la acción de mejorar el requerimiento de software y referencia la expresión a ser examinada.

| User Prompt |
|---|
| Indicate how the following software requirement can be improved. Software Requirement: "The system must allow the inventory manager to generate a list of missing products." |
| **Model** |
| The given software requirement can be improved … **Issues in the Original Requirement:** 1. **Ambiguity in "missing products"** – Does this mean out-of-stock items… **Improved Requirement:** *"The system must enable the inventory manager to generate a report listing products that are out of stock or below the predefined threshold. The report should be available in CSV and PDF formats and accessible via the inventory management dashboard. The inventory manager must be able to generate this report on demand and receive an automatic daily summary via email."*… |

**Fig. 5.** Resumen de la consulta genérica sobre el primer requerimiento.

Como puede observarse en la salida, el modelo propone un requerimiento mejorado. Esta expresión es luego evaluada contra la obtenida con la implementación simplificada del marco propuesto.

### 4.4 Evaluación de las expresiones con el marco propuesto

En esta sección de desarrolla una Prueba Conceptual utilizando ChatGPT para evaluar la factibilidad del enfoque. Con el objetivo de simplificar el ejemplo se utiliza ChatGPT como único modelo en las distintas etapas. En esta implementación del marco son tres los modelos necesarios, a) modelo evaluador de la expresión, b) modelo generador de preguntas y c) modelo generador de una nueva expresión mejorada. La decisión respecto de la calidad o no calidad del requerimiento es tomada por el analista y las respuestas a las preguntas que alimentan el último modelo fueron generadas para estas pruebas. Esta prueba carece de un agente, puesto que el analista cumplirá ese rol orquestando el uso de los distintos componentes.

**Modelo Evaluador (a).** Al inicio, la propuesta exige definir claramente cómo se analizarán los requerimientos con el modelo que evalúa la calidad. Esto permite obtener resultados claros acerca de qué problemas presentan las expresiones. Debido a que estas pruebas se realizan mediante la interacción con un modelo generativo, la primera interacción define el contexto y comprueba que el modelo posee información de éste, en este caso la guía de INCOSE (Figura 6).

| User Prompt |
|---|
| Do you know the characteristics of well-formed requirements: "Necessary, Appropriate, Unambiguous, Complete, Singular, Feasible, Verifiable, Correct and Conforming" pub- |
| lished in "INCOSE Guide for writing requirements "? |
| **Model** |
| Yes! The **INCOSE (International Council on Systems Engineering) Guide for Writing Requirements** defines well-formed requirements using the following characteristics:<br>1. **Necessary** – The requirement must be essential to meet stakeholder needs …<br>… |

**Fig. 6.** Resumen de la primera interacción con el modelo.

Luego de esta interacción inicial, se procede a evaluar el requerimiento. El prompt está compuesto por: la tarea a realizar, el texto del requerimiento y, por último, define el formato deseado para la respuesta (Figura 7).
Con la salida obtenida en la última interacción se puede tomar la decisión sugerida en el marco, acerca de si el requerimiento tiene la calidad esperada o no. En este caso, la decisión es tomada por el analista, aunque el marco deja abierta esta decisión que podría ser tomada de forma automática implementando un agente.

| **User Prompt** | | |
|---|---|---|
| Verify that the requirement meets these characteristics.<br>Requirement: "The system must allow the inventory manager to generate a list of miss-ing products."<br>Your answer should be only a table with the columns [Feature Name, Feature Detail, Fulfilled (yes/no)] | | |
| **Model** | | |
| **Feature Name** | **Feature Detail** | **Fulfilled (yes/no)** |
| Necessary | Essential for inventory management | Yes |
| Appropriate | At the correct level for system requirements | Yes |
| Unambiguous | Clearly defines the function | No |
| Complete | Contains all necessary information | No |
| Singular | Addresses only one function | Yes |
| Feasible | Achievable with existing technology | Yes |
| Verifiable | Testable via inspection or system validation | No |
| Correct | Accurately describes the needed function | Yes |
| Conforming | Follows standard requirement guidelines | No |

**Fig. 7.** Salida del modelo evaluador para el primer requerimiento.

Del análisis del ejemplo de la Figura 7 se concluye que el requerimiento no tiene la calidad esperada debido a que, según la respuesta del modelo, tres características deseadas en la propuesta de INCOSE no se cumplen. Estas características son "No ambiguo", "Completo" y "Verificable". La característica "Conforme" se desestima en todas las evaluaciones, ya que estas pruebas no incluyen información acerca de los estándares de la organización para la escritura de requerimientos.

**Modelo Generador de Preguntas (b).** La segunda interacción con un modelo tiene como fin obtener preguntas cuyas respuestas mejoren la expresión en las características no cumplidas, según la interacción anterior. Para este propósito se confecciona un prompt que describe la tarea a realizar y el formato de la salida esperada (Figura 8).

| User Prompt | |
|---|---|
| Define questions whose answers help to comply with the missing characteristics: "unambiguous, complete and verifiable". Your answer should only be a table with the columns [Feature Name (unambiguous, complete, or verifiable), suggested questions] | |
| **Model** | |
| **Feature Name** | **Suggested Questions** |
| Unambiguous | What format should the generated list follow (e.g., PDF, CSV, on-screen display)?... |
| Complete | Should the system allow filtering or sorting the missing product list? … |
| Verifiable | What criteria will be used to confirm the list is correctly generated? |

**Fig. 8.** Salida resumida del modelo generador de preguntas para el primer requerimiento.

**Modelo Generador de Requerimientos (c).** A partir de las preguntas generadas en la última interacción, se selecciona un conjunto que cubra todos los aspectos a mejorar y se generan respuestas sintéticas con el fin de culminar la aplicación del marco. Se confecciona un prompt como entrada al último modelo, que describe la tarea a realizar e incluye el texto original del requerimiento, las preguntas y respuestas para mejorarlo y los formatos propuestos para la redacción de requerimientos (Figura 9).

Como puede observarse en la salida que muestra la Figura 9, el modelo propone un requerimiento mejorado. Esta expresión incluye detalles extraídos por el modelo de las respuestas provistas.

| User Prompt |
|---|
| Improve the requirement using the answers to questions. The improved requirement must be expressed according to the Proposed Format. |
| Requirement: "The system must allow the inventory manager to generate a list of missing products." |
| Questions and answers: |
| Q1-What format should the generated list follow (e.g., PDF, CSV, on-screen display)? |
| A1-The list should be available in PDF and CSV formats, as well as an on-screen display for quick review. |
| Q2-… |
| Proposed formats: |
| F1- The <entity> shall <action verb> <object> <measurable outcome>. |
| F2- When <condition clause>, the <subject clause> shall <action verb clause> <object clause> <qualifying clause>. |
| **Model** |

> **Improved Requirement (Format F2):**
> *When the inventory manager or authorized personnel request a missing products list, the system shall generate a report that includes out-of-stock items and products below a predefined threshold, providing details such as product name, current quantity, supplier, and recommended reorder date. The list shall be available in PDF and CSV formats, as well as an on-screen display, and shall support filtering by category, supplier, and stock status, with sorting by priority.*

**Fig. 9.** Salida resumida del modelo generador de preguntas para el primer requerimiento.

### 4.5 Evaluación

Para evaluar las expresiones obtenidas con el marco y facilitar la comparación con las obtenidas según la propuesta de la sección 4.3, se utilizó el mismo prompt que representa el modelo evaluador (a) descripto en la Figura 2. Por cada requerimiento se evaluó la expresión original (RO), la generada con un prompt genérico (RG) y la generada con la implementación del marco (RM) (Tabla 3).

**Tabla 3.** Resultados de la evaluación sobre el requerimiento original (**RO**), el generado con el prompt genérico (**RG**) y el generado con la implementación del marco (**RM**).

| # | Requerimiento 1 | | | Requerimiento 2 | | | | Requerimiento 3 | | |
|---|---|---|---|---|---|---|---|---|---|---|
| **Característica** | RO | RG | RM | RO | RG | RM | RM | RO | RG | RM |
| Necesario | Si | Si | Si | Si | Si | **Si** | **Si** | Si | Si | Si |
| Apropiado | Si | Si | Si | No | Si | **Si** | **Si** | Si | Si | Si |
| No ambiguo | No | No | Si | No | Si | **Si** | **Si** | No | No | Si |
| Completo | No | No | Si | No | Si | **Si** | **Si** | No | No | Si |
| Singular | Si | No | No | No | No | **Si** | **Si** | Si | No | Si |
| Factible | Si | Si | Si | Si | Si | **Si** | **Si** | Si | Si | Si |
| Verificable | No | Si | Si | No | Si | **Si** | **Si** | Si | No | Si |
| Correcto | Si | Si | Si | No | Si | **Si** | **Si** | Si | Si | Si |
| % Caract. Cumplidas | 62.5 | 62.5 | **87.5** | 25 | 87.5 | **100** | **100** | 75 | 50 | **100** |

Para interpretar la Tabla 3, se aclara que para el Requerimiento 2 existe una columna extra en la expresión obtenida con el marco propuesto. Esto se debe a que la propuesta detectó la no singularidad y generó dos expresiones para solucionarla.

Los resultados pueden analizarse de diferentes formas. En principio, se pretende medir la "calidad" de las expresiones según la evaluación devuelta por el modelo. Para esto, se plantea una métrica calculada como el porcentaje de características cumplidas por cada expresión de requerimiento. Así, se obtiene una medida que se visualiza en la última fila de la Tabla 3. Claramente, en las tres expresiones de prueba, los mejores resultados son obtenidos con el marco propuesto con tres casos, en los que logra el 100% para esta métrica (Figura 10).

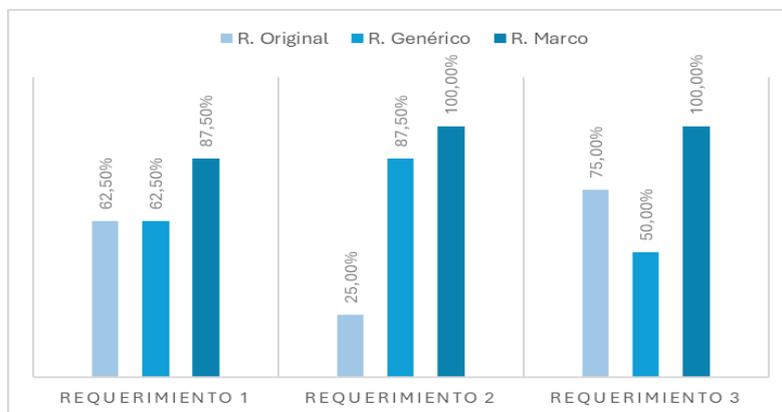

**Fig. 10.** Comparación, según la métrica propuesta, de la calidad de los distintos requerimientos.

## 5    Conclusiones y Trabajos Futuros

El framework propuesto tiene un gran potencial, sin embargo, es esencial reconocer que sólo se presenta una prueba conceptual. La implementación utilizada en este trabajo es sencilla y hace uso de sólo tres modelos representados por interacciones con ChatGPT. Sin embargo, estas pruebas preliminares sirven para recoger los primeros resultados y conclusiones, así como también guiar futuros trabajos.

La estrategia de dividir el problema en tareas simples y pequeños incrementos a la solución, aporta a los modelos el conocimiento y el contexto que se va acumulando luego de cada interacción. Esto no es menor, ya que ese contexto generado en los pasos previos suma información a la entrada de la interacción que permite una respuesta más precisa y completa. Esto puede observarse, por ejemplo, en el Requerimiento 2, donde primero se detecta la no singularidad, luego se genera una pregunta acerca de la posibilidad de dividir la expresión y, por último, ante una respuesta afirmativa a esta pregunta, el modelo genera dos expresiones.

Igualmente, es de destacar cómo esta implementación simplificada del marco pudo mejorar las expresiones hasta lograr un 96.78% en la métrica propuesta, promediando las medidas logradas para las cuatro expresiones generadas.

Además, se puede observar que las expresiones generadas por una consulta genérica no mejoran en todos los casos la expresión original, empeorándola a veces como en el requerimiento 3. Estas expresiones logran un puntaje de 66.6% con la métrica propuesta y promediando los puntajes. Puede observarse, por ejemplo, que la consulta genérica no soluciona la no singularidad e inclusive, a veces, la causa.

Por último, se destaca cómo el último modelo generador de expresiones de requerimientos puede crearlas siguiendo patrones que son anexados al prompt de generación.

Son varios los posibles trabajos futuros que se desprenden de la propuesta, como ser la implementación de un prototipo de software funcional que implemente un agente para automatizar el proceso y así permitir su aplicación con datos reales de proyectos de desarrollo de software. Esto permitiría evaluar la propuesta a mayor escala y con distintas implementaciones del marco.

La variedad de modelos y artefactos posibles exige una mejor y más completa definición de la propuesta, y una evaluación sobre qué interfaces y repositorios se pueden utilizar para conectar los modelos, además de cómo implementar la vectorización de texto en la base de datos -en el caso de utilizar RAG- o cómo entrenar modelos y preparar los conjuntos de datos son algunos de los desafíos.

## Referencias